\documentclass[nofootinbib,aps,10pt]{revtex4-1}

\pdfoutput=1

\usepackage{graphicx,color,verbatim,acronym,epsfig,subfigure,slashed,skak,geometry}
\usepackage{latexsym,amsfonts,amssymb,amsmath,makeidx,mathrsfs,multirow,lineno,bm,bbm,ifpdf}
\usepackage{hyperref}
\hypersetup{
    colorlinks,
    linkcolor={black},
    citecolor={black},
    urlcolor={black}
}

\usepackage[utf8]{inputenc}
\allowdisplaybreaks

\newcommand{\GeV}      {~\mathrm{GeV}}
\newcommand{\TeV}      {~\mathrm{TeV}}

\def\beqn{\begin{eqnarray}}
\def\eeqn{\end{eqnarray}}
\def\beqs{\begin{subequations}}
\def\eeqs{\end{subequations}}
\def\beq{\begin{equation}}
\def\eeq{\end{equation}}
\def\ba{\begin{array}}
\def\ea{\end{array}}

\def\non{\nonumber\\}
\def\hf{\frac{1}{2}}

\def\gU{\rm U}
\def\gSU{\rm SU}

\def\mL{\mathcal{L}}

\usepackage[normalem]{ulem}

\begin{document}
\title{\Large Further study of the global minimum constraint on the two-Higgs-doublet models: LHC searches for heavy Higgs bosons}
\bigskip

\author{Ning Chen$^{1}$}
\email{chenning$\_$symmetry@nankai.edu.cn}
\author{Chun Du~$^{2}$}
\email{duchun@utibet.edu.cn}
\author{Yongcheng Wu~$^{3}$}
\email{ycwu@physics.carleton.ca}
\author{Xun-Jie Xu~$^{4}$}
\email{xunjie@mpi-hd.mpg.de}
\affiliation{
	$^1$ School of Physics, Nankai University, Tianjin 300071, China
	\\
	$^2$ Key Laboratory of Cosmic Rays (Tibet University), Ministry of Education, Lhasa 850000, Tibet, China
	\\
	$^3$ Ottawa-Carleton Institute for Physics, Carleton University, 1125 Colonel By Drive, Ottawa, Ontario K1S 5B6, Canada 
	\\
	$^4$ 
	Max-Planck-Institut f\"ur Kernphysik, Postfach 103980, D-69029 Heidelberg, Germany
}
\date{\today}

\begin{abstract}
The usually considered vacuum of the two-Higgs-doublet model (2HDM) could be unstable if it locates at a local but not global minimum (GM) of the scalar potential. 
By requiring the vacuum to be a GM, we obtain an additional constraint, namely the GM constraint, on the scalar potential. 
In this work, we explore the GM constraint on the $CP$-conserving general 2HDM.
This constraint is found to put limits on the soft $\mathbb{Z}_2$ breaking mass parameter $m_{12}^2$ and also squeeze the heavy $CP$-even Higgs boson mass into larger values for the $m_{12}^2< 0$ case.
Combined with the current global signal fits from the LHC measurements of the $125\,\GeV$ Higgs boson, we discuss the phenomenological implications for the heavy Higgs boson searches at the LHC.
\end{abstract}

\maketitle
\baselineskip=16pt

\pagenumbering{arabic}

\vspace{1.0cm}
\tableofcontents

\newpage


\section{Introduction}
\label{section:intro}

In the studies of new physics beyond the Standard Model (BSM), it is quite often that one has an extended Higgs sector. 
A simple and well-known example is the two-Higgs-doublet model (2HDM),~\footnote{See Ref.~\cite{Branco:2011iw} for a comprehensive review. } which was motivated from several different aspects, such as supersymmetry~\cite{Haber:1984rc,Djouadi:2005gj}, $CP$ violation~\cite{Lee:1973iz}, and axion models~\cite{Kim:1986ax}.
With an additional Higgs doublet introduced, the Higgs potential in the 2HDM may develop several different minima. 
Therefore, one may encounter the possibilities as follows: (i) one Higgs doublet does not acquire a vacuum expectation value (VEV), (ii) the Higgs VEVs break the $CP$ symmetry, or (iii) the Higgs VEVs even break the $\gU(1)_{\rm EM}$ symmetry.
It has previously been studied in Refs.~\cite{Ferreira:2004yd,Barroso:2005sm,Barroso:2005tq,Barroso:2005da,Ivanov:2006yq,Barroso:2006pa,Ivanov:2007de,Barroso:2007rr,Ivanov:2007ja,Ivanov:2008er,Ivanov:2010ww,BarroseSa:2009ak,Ginzburg:2009dp,Ivanov:2010wz,Battye:2011jj,Barroso:2012mj,Barroso:2013ica,Barroso:2013awa,Barroso:2013kqa,Ivanov:2015nea} that several minima can coexist in the 2HDM potential so that the desired vacuum might be a local minimum that could decay into a deeper one through quantum tunneling~\cite{Coleman:1977py, Callan:1977pt}, causing instability of the desired vacuum.

To avoid the vacuum instability, one may impose a global minimum (GM) condition for the desired vacuum.
This leads to new constraints on the Higgs potential, in addition to the conventional bounded-from-below (BFB) constraints and the unitarity bounds.
Recently, the GM condition of the 2HDM potential has been analytically formulated in Ref.~\cite{Xu:2017vpq} and tentatively applied to constrain the general 2HDM. 
It has been demonstrated that the GM constraint can sometimes be robust in constraining the parameter space of the 2HDM.
\footnote{Typically in many BSM models with scalar extensions, the GM conditions of the desired vacua can be nontrivial and deserve further studies. The GM conditions of some models have been studied before, such as the Georgi-Machacek model in Ref.~\cite{Hartling:2014zca},  the Type II Seesaw model in Ref.~\cite{Xu:2016klg}, and the left-right symmetric model in Ref.~\cite{Dev:2018foq}. 
}

In this work, we further study the GM constraint on the 2HDM, with the focus on the phenomenological implications at the LHC.
It turns out that the GM condition is likely to put constraints on the masses of heavy Higgs bosons and the soft $\mathbb{Z}_2$ breaking scale of $m_{12}^2$ in the 2HDM.
In turn, these constraints are directly connected to the Higgs self-couplings in the 2HDM. 
From the experimental point of view, the Higgs self-couplings are likely to be probed by the high-luminosity (HL) and/or high-energy (HE) LHC runs, by looking for the Higgs boson pair productions.
Since the discovery of the $125\,\GeV$ Higgs boson at the LHC, a lot of efforts have been made in probing such processes in different new physics models at the LHC~\cite{Baglio:2012np, Shao:2013bz, Chen:2013emb, Chen:2014xra, Barger:2014qva, Bian:2016awe, Cao:2016zob, Kilian:2017nio, Cacciapaglia:2017gzh, DiLuzio:2017tfn, Alves:2017ued, Corbett:2017ieo, Grober:2017gut, Ren:2017jbg, Basler:2017uxn,Dawson:2017jja,Adhikary:2017jtu, Goncalves:2018qas}.
Since the Higgs self-couplings in the 2HDM can determine the corresponding partial decay widths of a heavy Higgs boson into lighter Higgs pairs, the future experimental searches for heavy Higgs bosons in the 2HDM may also be sensitive to the GM constraint.

The layout of this paper is described as follows. 
In Sec.~\ref{section:min}, we revisit the $CP$-conserving general 2HDM, where we put emphasis on the GM constraint on the 2HDM potential.
This constraint, together with the usual tree-level BFB and perturbative unitarity constraints, will be imposed on the 2HDM parameter space. 
In Sec.~\ref{sec:masses}, we consider benchmark models in two different scenarios, namely, the degenerate heavy Higgs boson scenario of $M_A=M_H=M_\pm$, and the heavy Higgs boson spectrum involving exotic decays. 
It turns out that the GM condition leads to additional restrictions on the parameter space.
In Sec.~\ref{section:Hpair}, we study the LHC phenomenologies based on the GM constraints on the benchmark models.
Since the GM constraint on the $m_{12}^2$ parameter will control the Higgs boson self-couplings in the 2HDM, the pair productions of both SM-like and BSM Higgs bosons at the LHC can be relevant to this constraint.
The current LHC $13\,\TeV$ searches for the Higgs boson pairs, as well as other exotic heavy Higgs boson decay modes are imposed to the benchmark models with the GM constraint taken into account.
Finally, we conclude in Sec.~\ref{section:conclusion}.


\section{The general 2HDM and the GM constraint}
\label{section:min}

\subsection{The general 2HDM}

The scalar potential of the general 2HDM is written as follows
\beqn\label{eq:V2HDM}
V(\Phi_1\,,\Phi_2)&=&m_{11}^2|\Phi_1|^2+m_{22}^2|\Phi_2|^2- m_{12}^2 (  \Phi_1^\dag\Phi_2+{\rm H.c.})\non
&&+\hf\lambda_1(\Phi_1^\dag\Phi_1)^2+\hf\lambda_2(\Phi_2^\dag \Phi_2)^2+\lambda_3|\Phi_1|^2 |\Phi_2|^2+\lambda_4 |\Phi_1^\dag \Phi_2|^2\non
&&+\hf  \lambda_5 \Big[ (\Phi_1^\dag\Phi_2)^2+{\rm H.c.}  \Big]\,,
\eeqn
where all the couplings are real for the $CP$-conserving case.
Here, we do not include the $\mathbb{Z}_2$ broken terms of $\lambda_6 (\Phi_1^\dag\Phi_1) (\Phi_1^\dag\Phi_2)+\lambda_7 (\Phi_2^\dag\Phi_2) (\Phi_1^\dag\Phi_2) + {\rm H.c.}$, focusing our study on the potential with softly broken $\mathbb{Z}_2$ symmetry.

The potential in Eq.~\eqref{eq:V2HDM} contains eight parameters, namely  $m_{11}^2$, $m_{12}^2$, $m_{22}^2$, and $\lambda_{1\cdots5}$, which are usually referred to as the parameters in the generic basis. In phenomenological studies, it is usually more convenient to work in the so-called physical basis, including the five physical boson masses of $(M_h\,, M_H\,, M_A\,, M_\pm)$, two mixing angles of $(\alpha\,,\beta)$, a soft $\mathbb{Z}_2$ broken mass squared term of $m_{12}^2$, and the electroweak VEV $v\approx246 \GeV$.

In the general 2HDM,  there could be tree-level flavor-changing neutral currents (FCNC), which are well-known constraints on such model.
To alleviate the tree-level FCNC process constraints, the SM fermions of a given representation are usually assigned to a single Higgs doublet. 
We focus on the so-called Type I and Type II Yukawa couplings of
\beqn
\mL&\supset& \sum_{h_i=h\,,H}  - \frac{m_f}{v} \left( \xi_i^f \bar f f   h_i + \xi_A^f \bar f i \gamma_5 f A \right) \,,
\eeqn
with
\beqs\label{eqs:Higgs_coup}
\beqn
\textrm{Type I}&:& \xi_h^f= \sin(\beta-\alpha) + \frac{\cos(\beta-\alpha)}{\tan\beta} \,,\qquad \xi_{H}^f=\cos(\beta-\alpha) - \frac{ \sin(\beta-\alpha) }{\tan\beta}\non
&& \xi_A^u= \frac{1}{\tan\beta}\,, \qquad \xi_A^{d\,, \ell}= - \frac{1}{\tan\beta}\,,\\
\textrm{Type II}&:&\xi_h^u=\sin(\beta-\alpha) + \frac{ \cos(\beta-\alpha) }{\tan\beta}\,,\qquad \xi_h^{d\,,\ell}= \sin(\beta-\alpha) - \cos(\beta-\alpha)\,\tan\beta \,,\non
&&\xi_H^u=\cos(\beta-\alpha)- \frac{ \sin(\beta-\alpha)}{\tan\beta}\,,\qquad \xi_H^{d\,,\ell}= \cos(\beta-\alpha) + \sin(\beta-\alpha)\,\tan\beta\,,\non
&& \xi_A^u= \frac{1}{\tan\beta}\,,\qquad \xi_A^{d\,,\ell} = \tan\beta\,.
\eeqn
\eeqs
Besides, two $CP$-even Higgs bosons couple to the gauge bosons such that
\beqn
\mL&\supset& \sum_{h_i=h\,,H}  a_i \left( 2  \frac{m_W^2}{v} W_\mu^+ W^{-\,\mu} +  \frac{m_Z^2}{v} Z_\mu Z^\mu \right)  h_i\,,
\eeqn
with 
\beqn
&&a_h= \sin(\beta-\alpha)\,,\qquad a_H = \cos(\beta-\alpha)\,.
\eeqn

\begin{figure}[tb]
\centering
\includegraphics[width=0.45\textwidth]{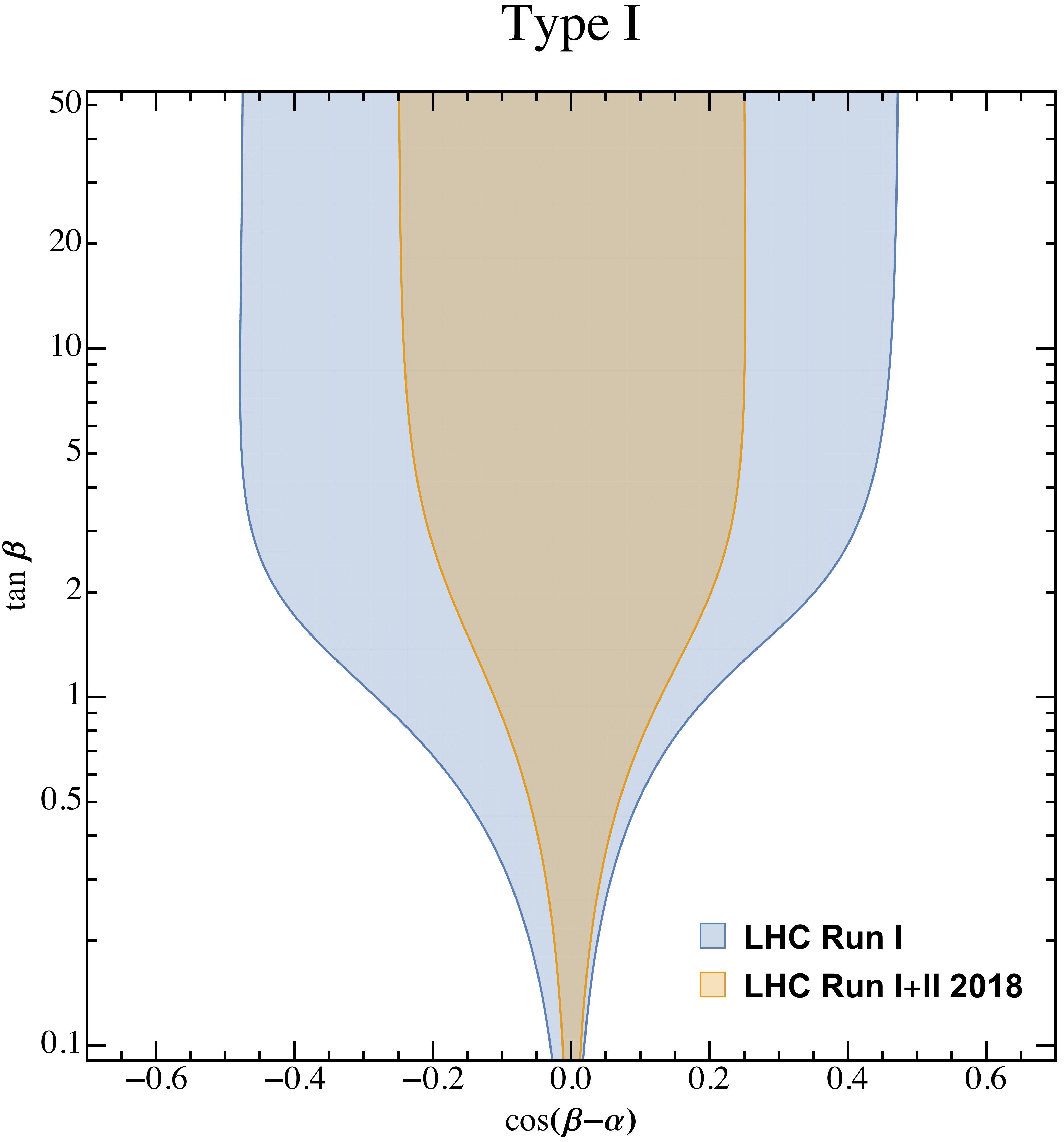}
\includegraphics[width=0.45\textwidth]{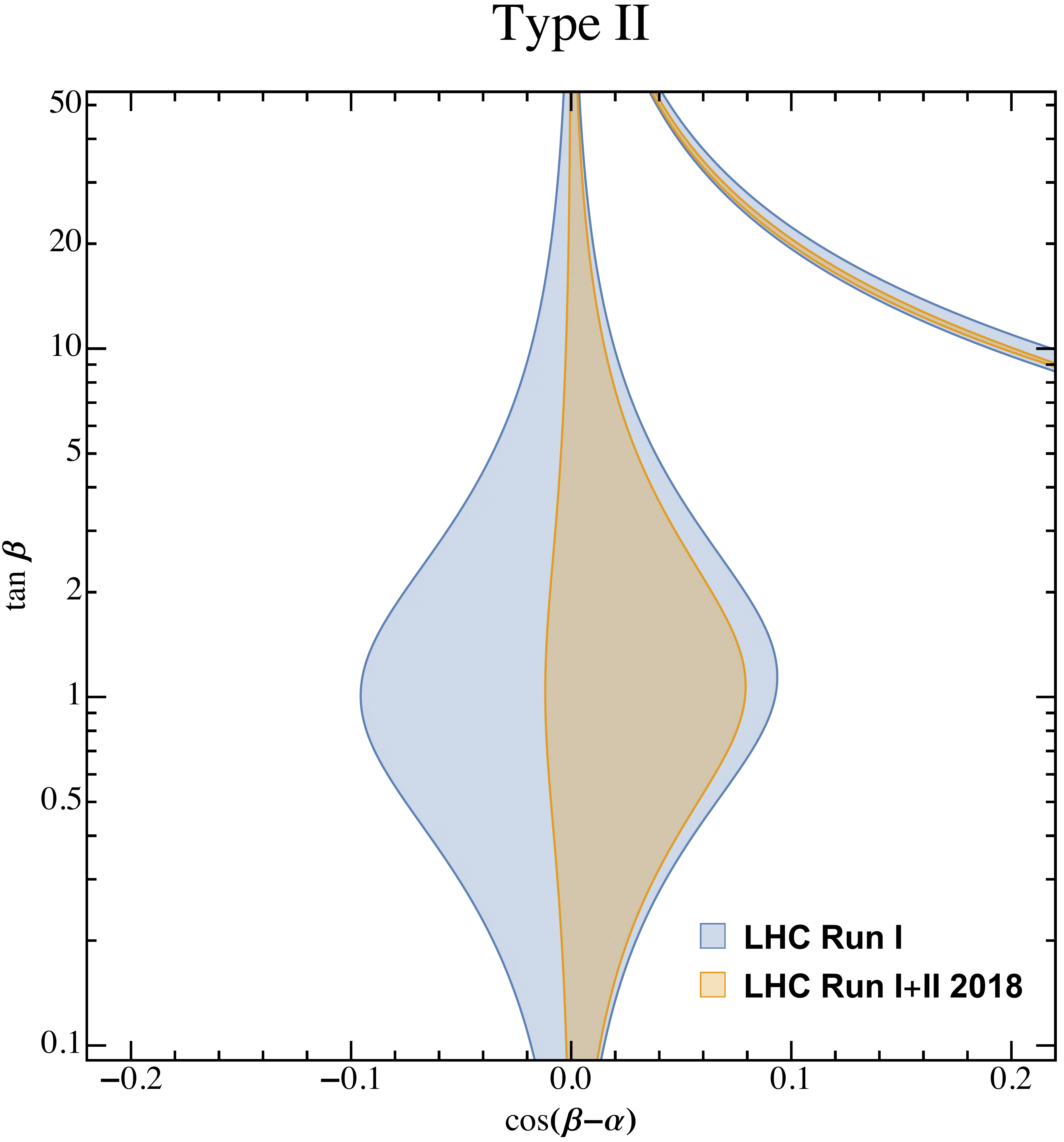}
\caption{
The combined LHC run-I and run-II constraints on the $125$ GeV Higgs boson signal strengths in terms of 2HDM parameters $\cos(\beta-\alpha)$ and $\tan\beta$.
Both Type I (left) and Type II (right) cases are displayed.
}
\label{fig:THDM_hsig}
\end{figure}

The current LHC run I and run II have measured the signal strengths of the SM-like $125$ GeV Higgs boson\footnote{Throughout the context, we always assume that $M_h=125$ GeV, while all other Higgs bosons are heavier.} via different channels~\cite{Aad:2014eha, Aad:2014eva, ATLAS:2014aga, Aad:2015gra, Aad:2014xzb, Aad:2015vsa, Khachatryan:2014ira, Chatrchyan:2013mxa, Chatrchyan:2013iaa, Khachatryan:2015ila, Chatrchyan:2013zna, Chatrchyan:2014nva, ATLAS:2017myr, ATLAS:2017ovn, ATLAS:2017cju, ATLAS-CONF-2016-112, Aaboud:2017xsd,Campos:2017dgc, ATLAS-CONF-2016-080, Aaboud:2017ojs, CMS:2017jkd, Sirunyan:2017exp, CMS-PAS-HIG-16-021, Sirunyan:2017dgc, Sirunyan:2017khh, CMS:2017lgc,Khachatryan:2016vau,ATLAS-CONF-2018-031,Sirunyan:2018koj}.
Here, they are combined to obtain the $95\%$ C.L. regions in the $\cos(\beta-\alpha)$ vs $\tan\beta$ plane, as shown in Fig.~\ref{fig:THDM_hsig}.
In both Type I and Type II, the alignment limit of $\beta-\alpha=\pi/2$ is favored by the global fit. 
For the Type I 2HDM with $\tan\beta\gtrsim 2$, $| \cos(\beta-\alpha) |$ is constrained to be less than about $0.4$ with the LHC run-I and run-II data. 
This is envisioned to be further constrained to be less than $0.2$ with the HL-LHC runs in Ref.~\cite{Gu:2017ckc}.
For the Type II 2HDM, large/small $\tan\beta$ inputs will enhance the Yukawa couplings $\xi_h^{d\,,\ell}/\xi_h^u$.
Thus, the region around $\tan\beta=1$ accommodates the largest deviation from the alignment.
The current LHC run-I and run-II measurements constrain $\cos(\beta-\alpha)$ in the range of $( -0.01\,,0.08)$ approximately with $\tan\beta=1$ (except for the wrong-sign Yukawa coupling region~\cite{Ferreira:2014naa, Han:2017etg}).

\subsection{The GM constraint on the 2HDM potential\label{subsec:GM}}

\begin{table}
\centering
\caption{\label{tab:All-local-minima}
All possible local minima of the scalar potential. 
``$\times$''  denotes a nonzero component, and ``$*$'' stands for an arbitrary value (can be zero or nonzero). 
}
\begin{ruledtabular}

\begin{tabular}{cccccc}
 & \textcolor{white}{.}\hspace{0.8cm}Type A\hspace{0.8cm}\textcolor{white}{.} & \textcolor{white}{.}\hspace{0.8cm}Type B\hspace{0.8cm}\textcolor{white}{.} & \textcolor{white}{.}\hspace{0.8cm}Type C\hspace{0.8cm}\textcolor{white}{.} & \textcolor{white}{.}\hspace{0.8cm}Type D\hspace{0.8cm}\textcolor{white}{.} & \textcolor{white}{.}\hspace{0.8cm}Type E\hspace{0.8cm}\textcolor{white}{.}\tabularnewline
\hline 
\rule[-5ex]{0pt}{10ex}  $\langle\phi_{1}\rangle$, $\langle\phi_{2}\rangle$ & 
$\left[ \ba{c}  0 \\  \times \ea\right]$, $\left[\ba{c} \times \\ * \ea \right]$ 
& $\left[\ba{c} 0 \\ 0  \ea\right]$, $\left[\ba{c}
0\\  \times  \ea\right]$ 
& $\left[\ba{c}  0  \\  \times  \ea\right]$, $\left[\ba{c} 0  \\  0 \ea\right]$ 
& $\left[\ba{c}  0 \\  \times  \ea\right]$, $\left[\ba{c}  0 \\  \times \ea\right]$ 
& $\left[\ba{c}  0  \\  0  \ea\right]$, $\left[\ba{c}  0 \\  0 \ea\right]$\tabularnewline
\rule[-3ex]{0pt}{8ex}  $(q_{1}\,, q_{2}\,, z)$ & Eq.~\eqref{eq:TypeA_sol} & $(0\,,-\frac{m_{22}^{2}}{\lambda_{2}}\,,0)$ & $(-\frac{m_{11}^{2}}{\lambda_{1}}\,, 0\,,0)$ & Eq.~\eqref{eq:m11m22} & $(0\,,0\,,0)$\tabularnewline
\rule[-5ex]{0pt}{10ex}  $\ba{c}
{\rm Existence}\\
{\rm condition} \ea$ 
& $\ba{c} q_{1}\,, q_{2}>0 \\ |z|^{2}<q_{1}q_{2}  \ea $ 
& $q_{2}>0$ & $q_{1}>0$ 
& $\ba{c}  q_{1},\thinspace q_{2}>0 \\ |z|^{2}=q_{1}q_{2} \ea$ 
& $/$~\footnote{Not required.}\tabularnewline
\hline 
\rule[-3ex]{0pt}{8ex}  $V_{{\rm min}}$ & Eq.~\eqref{eq:TypeA_Vmin} & $-\frac{m_{22}^{4}}{2\lambda_{2}}$, Eq.~\eqref{eqs:VminB} & $-\frac{m_{11}^{4}}{2\lambda_{1}}$, Eq.~\eqref{eqs:VminC} & Eq.~\eqref{eqs:VminD}
& 0\tabularnewline
\end{tabular}

\end{ruledtabular}
\end{table}

As explained in the Introduction, to guarantee the absolute stability of the usually considered vacuum, we shall impose the GM constraint on the potential. 
First, we will present all the possible minima of the potential at the tree level and discuss the condition of the desired one being a global minimum. 
We realize that loop corrections can also have important influence on the GM constraint.
Subsequently, we will also address the issue of including loop corrections.

At the tree level, by defining the following three $\gSU(2)_L$ invariants of
\beqn
&&q_{1}\equiv | \Phi_{1}|^2=\Phi_1^\dag \Phi_1, \qquad q_{2}\equiv | \Phi_{2}|^2=\Phi_2^\dag \Phi_2,\qquad z\equiv \Phi_1^\dag \Phi_2= |z| e^{i\theta}\,,
\eeqn
%
%
the potential can be rewritten as
\beqn
V(\Phi_1\,,\Phi_2)&=&m_{11}^2 q_1 +m_{22}^2 q_2- ( m_{12}^2  z + {\rm H.c.} )\non
&&+\hf\lambda_1(q_1)^2 + \hf\lambda_2 (q_2)^2+\lambda_3 q_1 q_2 +\lambda_4 |z|^2+\hf  ( \lambda_5 z^2+{\rm H.c.}  ) \,.
\eeqn
In principle, one can directly minimize the above potential with respect to $q_1$, $q_2$, and $z$. However, one should notice that by definition, the three $\gSU(2)_L$ invariants of $(q_1\,, q_2\,, z)$ satisfy the boundary conditions of
\beqn\label{eq:boundary}
&& q_{1\,,2}\geq 0\,, \qquad q_1 q_2 \geq |z|^2\,.
\eeqn

Depending on whether the minima are on one of the boundaries in Eq.~\eqref{eq:boundary}, we can classify the minima into five types, namely,
\beqs
\beqn
\textrm{Type A}&:& q_1>0\,,~~ q_2 > 0 \,, ~~ q_1 q_2 > |z|^2\,,\\
\textrm{Type B}&:& q_1=0\,,~~ q_2 > 0 \,, ~~ z=0\,,\\
\textrm{Type C}&:& q_1>0\,,~~ q_2 = 0 \,, ~~ z=0 \,,\\
\textrm{Type D}&:& q_{1\,,2}>0\,,~~ q_1 q_2 = |z|^2 \,,\\
\textrm{Type E}&:& q_{1\,,2} = z =0\,,
\eeqn
\eeqs
where, e.g.,  Type A is not on any of the boundaries and Type E is on all of the boundaries.
All of the five types of minima have been solved in Ref.~\cite{Xu:2017vpq} and summarized in Table~\ref{tab:All-local-minima}.

The row in Table~\ref{tab:All-local-minima} containing the explicit forms of $\langle\phi_{1}\rangle$ and $\langle\phi_{2}\rangle$ indicates that Type D is the usually desired vacuum of 2HDM. 
Type A minima could break $\gU(1)_{\rm EM}$, while Type B and Type C minima appear in the so-called inert 2HDM. 
Type E is a trivial solution that is listed here for completeness.

The solution of Type A is given by
\beqn\label{eq:TypeA_sol}
\textrm{Type A}:\ 
\left( \ba{c}
q_1 \\ q_2 \\ z \\ z^* \ea   \right)&=& \Lambda^{-1}\, b\,,
\eeqn
where
\beqn
&&\Lambda=\left( \ba{cccc} 
\lambda_1 & \lambda_3 & 0 & 0 \\
\lambda_3 & \lambda_2 & 0 & 0 \\
0 & 0 & \lambda_5 & \lambda_4 \\
0 & 0 & \lambda_4 & \lambda_5^*    \ea \right)\,,\qquad  b= \left( \ba{c} 
-m_{11}^2 \\ - m_{22}^2 \\ m_{12}^2 \\  (m_{12}^2 )^*  \ea \right) \,.
\eeqn
And the corresponding potential minimum is
\beqn\label{eq:TypeA_Vmin}
V_{\rm min\,,A}&=& - \hf b^T \Lambda^{-1} b = \frac{ - m_{11}^4 \lambda_2 - m_{22}^4 \lambda_1 + 2 m_{11}^2 m_{22}^2 \lambda_3 }{2 ( \lambda_1 \lambda_2 - \lambda_3^2 )}- \frac{   (m_{12}^2)^2 }{ \lambda_4+ \lambda_5}\,.
\eeqn

The Type D minimum is determined by $\partial V/\partial q_{1\,,2}= 0$:
\beqs\label{eq:m11m22}
\beqn
m_{11}^2 &=& m_{12}^2  \tan\beta  - \Big[ \lambda_1 \cos^2\beta +  (\lambda_{3}+\lambda_{4}+\lambda_{5})  \sin^2\beta \Big] q  \,,\label{eq:m112}\\
m_{22}^2 &=& m_{12}^2/ \tan\beta - \Big[  \lambda_2 \sin^2\beta + (\lambda_{3}+\lambda_{4}+\lambda_{5}) \cos^2\beta  \Big] q  \,,\label{eq:m222}
\eeqn
\eeqs
where $q \equiv q_1 + q_2$. 
Given the potential parameters of ($m_{11}^2$, $\lambda_{1}$, $\lambda_{2}$, ...), Eqs.~\eqref{eq:m112} and \eqref{eq:m222}
can be solved with respect to $q$ and $\beta$, which can be further converted to $q_1$ and $q_2$ according to $q_2/q_1=\tan^2\beta$ and $q_1+q_2=q$. 
In practical use with physical inputs, Eqs.~\eqref{eq:m112} and \eqref{eq:m222} are commonly used to evaluate $m_{11}^2$ and $m_{22}^2$ for given $\tan \beta$ and $v$, together with the quartic couplings determined by
%
%
%
\beqs\label{eqs:lambda_physical}
\beqn
\lambda_1&=& \frac{ M_h^2 \sin^2\alpha + M_H^2 \cos^2\alpha - m_{12}^2 \tan\beta }{ v^2\, \cos^2\beta}\,,\\
\lambda_2&=& \frac{ M_h^2 \cos^2\alpha + M_H^2 \sin^2\alpha - m_{12}^2 /\tan\beta }{ v^2\, \sin^2\beta}\,,\\
\lambda_3&=& \frac{1}{v^2} \Big[   \frac{ ( M_H^2 - M_h^2 ) \sin\alpha \cos\alpha }{ \sin\beta \cos\beta} + 2 M_\pm^2 - \frac{ m_{12}^2 }{\sin\beta \cos\beta } \Big]\,,\\
\lambda_4&=& \frac{1}{v^2} ( M_A^2 - 2 M_\pm^2 + \frac{ m_{12}^2 }{\sin\beta \cos\beta} )\,,\\
\lambda_5&=&  \frac{1}{v^2}  ( \frac{ m_{12}^2 }{\sin\beta \cos\beta}  - M_A^2 )\,.
\eeqn
\eeqs
The potential minima for the Type B, Type C, and Type D cases can be expressed as follows in the physical basis:
\beqs\label{eqs:VminBCD}
\beqn
V_{\rm min\,,B}&=&-\frac{v^2 \cos^2\beta \Big[  (M_H^2-M_h^2)\sin\alpha \cos\alpha +(M_H^2 \sin^2 \alpha + M_h^2 \cos^2\alpha  )\tan\beta - 2m_{12}^2  \Big]^2}{8  \left(M_h^2 \cos^2\alpha+M_H^2 \sin^2\alpha - m_{12}^2/\tan\beta \right)} \,, \label{eqs:VminB}\\
V_{\rm min\,,C}&=&-\frac{v^2 \sin^2\beta \Big[ (M_H^2-M_h^2) \sin\alpha \cos\alpha + (M_H^2 \cos^2\alpha + M_h^2 \sin^2\alpha )/\tan\beta-2m_{12}^2  \Big]^2}{8 \left( M_H^2 \cos^2\alpha +M_h^2 \sin^2\alpha-m_{12}^2 \tan\beta \right)} \,,\label{eqs:VminC}\\
V_{\rm min\,,D}&=&-\frac{v^2}{16} \Big[ M_H^2+ M_h^2+(M_H^2-M_h^2) \cos(2\alpha-2\beta)  \Big]\,.\label{eqs:VminD}
\eeqn
\eeqs
Apparently, the minimal values of the 2HDM potential in the Type B and Type C cases are essentially controlled by the input parameters of $(M_h\,,M_H\,,m_{12}^2\,, \alpha\,,\beta)$, while the minimal value in the Type D case is independent of $m_{12}^2$.

The GM constraint requires that the Type D minimum is a GM of the potential in order to protect the corresponding vacuum from decaying to other vacua.
To infer whether it is a GM, one can compute all the possible minima listed in Table~\ref{tab:All-local-minima} and then compare their $V_{\rm min}$'s. 
It is important to mention that the possible minima listed in Table~\ref{tab:All-local-minima} do not necessarily exist. Table~\ref{tab:All-local-minima} only provides the possible solutions of the first derivatives vanishing, which should be further checked by the existence conditions in Table~\ref{tab:All-local-minima}. 
If $(q_1,\, q_2,\, z)$ computed for a specific type violates the corresponding existence condition, the solution of this type does not exist. 
Otherwise, the solution exists. 
However, this does not necessarily imply it is a minimum since it could also be a maximum or saddle point. 
Technically, we do not need to check whether the obtained solutions are local minima or other extrema because we are only concerned about the Type D minimum. 
As long as $V_{\min}$ of Type D is lower than the potential values of other existing solutions, Type D must be a GM. 
Checking the existence of other types of solutions, however, is necessary in this procedure.

Note that the minima summarized in Table~\ref{tab:All-local-minima} are only for the tree-level potential, while at the loop level they may receive important corrections.
To include loop corrections, we use the package {\tt Vevacious}~\cite{Camargo-Molina:2013qva} which is capable of finding the minima of the one-loop effective potential given the minima of the tree-level potential. 
We will show that loop corrections can change the GM constraint quantitatively but not qualitatively. 
Therefore the tree-level analytic expressions can be useful tools for understanding the more complicated, loop-corrected GM constraint. Nevertheless, the loop corrections should be included for quantitative studies.

When applying the GM constraint, we shall first impose the BFB  \cite{Maniatis:2006fs, Ivanov:2006yq, Nishi:2007nh, Ferreira:2009jb, Ivanov:2018jmz} and perturbative unitarity bounds \cite{Arhrib:2000is, Kanemura:2015ska,Goodsell:2018tti,Goodsell:2018fex,Krauss:2018orw}.  
This is because the former is the premise of studying global minima, and the latter avoids too large 
quartic couplings. 
Although the unitarity bound is innocuous for the tree-level GM studies, it would drastically enhance the loop corrections. 
The BFB conditions of the tree-level potential are given as
\beqs\label{eqs:BFB}
\beqn
&& \lambda_{1\,,2} \geq 0\,,\\
&&\lambda_3 \geq - \sqrt{\lambda_1 \lambda_2}\,,\\
&& \lambda_3 + \lambda_4 - | \lambda_5 | \geq - \sqrt{\lambda_1 \lambda_2}\,.
\eeqn
\eeqs
However, as recently shown in Ref.~\cite{Staub:2017ktc}, the BFB constraints can be alleviated
with more feasible parameter regions when the radiative corrections
are taken into account. Therefore when studying the loop-level vacua
we should check the BFB status of the one-loop effective potential instead of the tree-level potential. 
In addition, the perturbative unitarity bounds, defined by the requirement that all the scalar scattering
amplitudes respect the unitarity condition, are usually considered in the high energy limit of $s\rightarrow\infty$. 
Recently it has been pointed out in Refs.~\cite{Goodsell:2018tti,Goodsell:2018fex,Krauss:2018orw} that some amplitudes at
finite $s$ might be significantly larger than that in the $s\rightarrow\infty$
limit. Hence, we will adopt the unitarity bounds improved by taking the
$s$ dependence into account, which are readily applicable using the
{\tt  SARAH} package \cite{Staub:2008uz,Staub:2013tta,Goodsell:2018tti}.


\section{The GM constraints  on some benchmarks \label{sec:masses}}

Given inputs of ($M_h$, $M_H$, $M_A$, $M_\pm$, $m_{12}^2$, $\alpha$, $\beta$, and $v$) in the physical basis , we can convert them to the potential parameters of ($m_{11}^2$, $m_{12}^2$, $m_{22}^2$, and $\lambda_{1\cdots5}$) in the generic basis and, with the method introduced in Sec.~\ref{subsec:GM}, infer whether the corresponding potential violates the GM condition.
In this section, we study the GM constraints on the parameters in the physical basis, focusing on two simple yet illustrative scenarios below:

\begin{itemize}
\item (i) all the heavy Higgs bosons are mass degenerate, i.e.,  $M_H = M_A = M_\pm$;
\item (ii) two of the heavy Higgs bosons are mass degenerate while the remaining are heavier or lighter than the degenerate mass\,---\,see Table~\ref{tab:2HDMexotic}. 
Such a mass spectrum allows exotic decays~\cite{Kling:2016opi,Coleppa:2017lue}.
\end{itemize}

\begin{figure}[tb]
\centering
\includegraphics[width=0.9\textwidth]{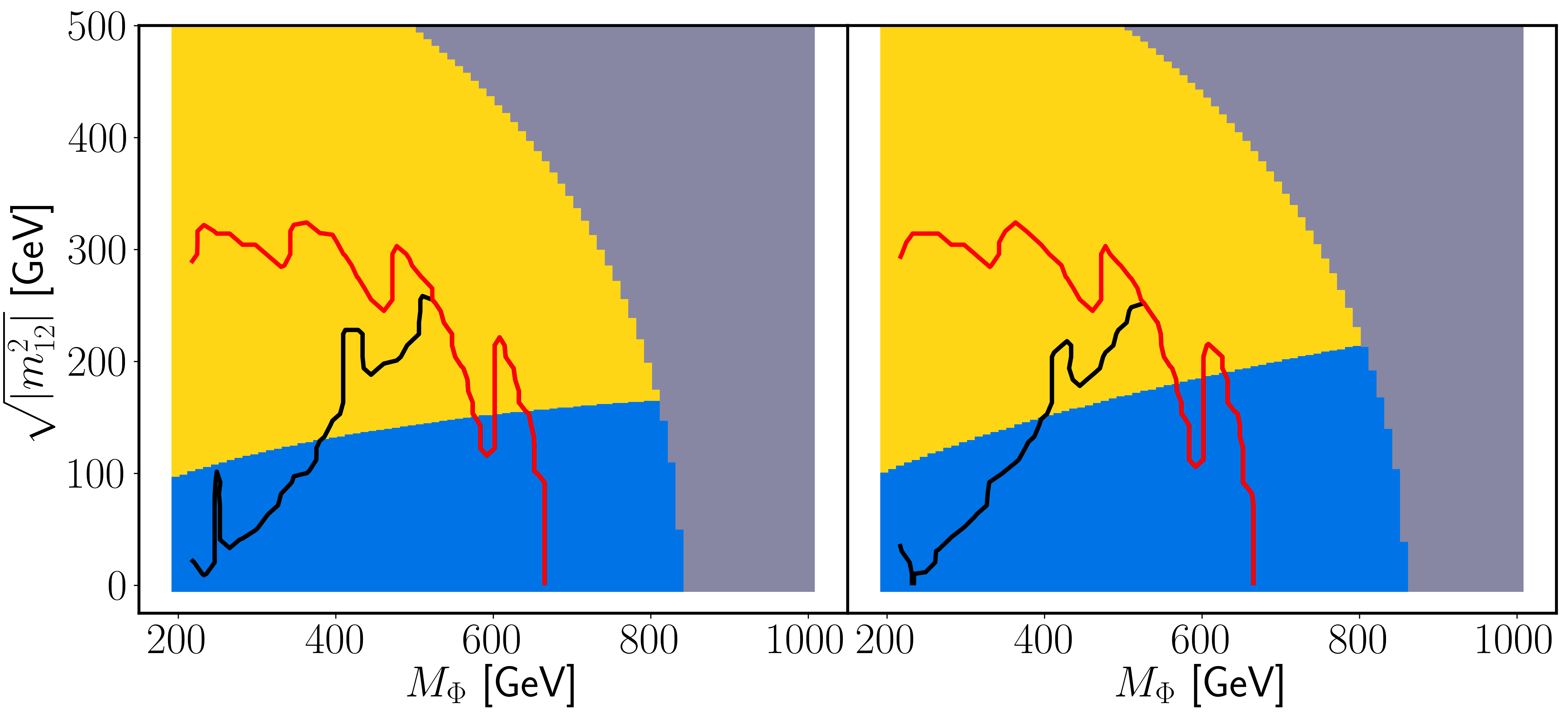}
\caption{
The GM constraints in the $( M_{\Phi}\,, \sqrt{ |m_{12}^2| })$ plane, where $m_{\Phi}$ is defined as the degenerate heavy Higgs boson mass $ M_H= M_A= M_\pm\equiv M_\Phi$.
The gray regions have already been excluded by the simplistic unitarity and BFB bounds, and the tree-level GM constraints further exclude the yellow regions, leaving only the blue regions allowed by all these constraints. When the loop corrections and the $s$ dependence are included, the unitarity and BFB bounds shift to the red curves, and the GM boundaries shift to the black curves.
The left and the  right panels  assume Type I 2HDM with $\cos(\beta-\alpha)=0.1$ and Type II 2HDM with $\cos(\beta-\alpha)=0.01$, respectively. 
}
\label{fig:THDM_constraints_degenerate}
\end{figure}

\begin{table}[htp]
\begin{center}
\begin{tabular}{|c|c|c|}
\hline
  & Mass planes & Decays    \\
\hline\hline
BP-1 & $M_A > M_H = M_\pm$ & $A\to (H^\pm W^\mp\,, HZ)$    \\
BP-2 & $M_A < M_H = M_\pm$ & $H\to (AZ\,, AA)\,, H^\pm\to  AW^\pm$   \\
BP-3 & $M_H > M_A = M_\pm$ & $H\to (AZ\,, AA)\,, H\to (H^+ H^-\,, H^\pm W^\mp)$   \\
BP-4 & $M_H < M_A = M_\pm$ & $A\to HZ\,, H^\pm \to H W^\pm$   \\
\hline\hline
\end{tabular}
\end{center}
\caption{Summary table of the different benchmark planes (BP) with exotic heavy Higgs boson decays in the 2HDM.}
\label{tab:2HDMexotic}
\end{table}%

\begin{figure}[tb]
\centering
\includegraphics[width=0.9\textwidth]{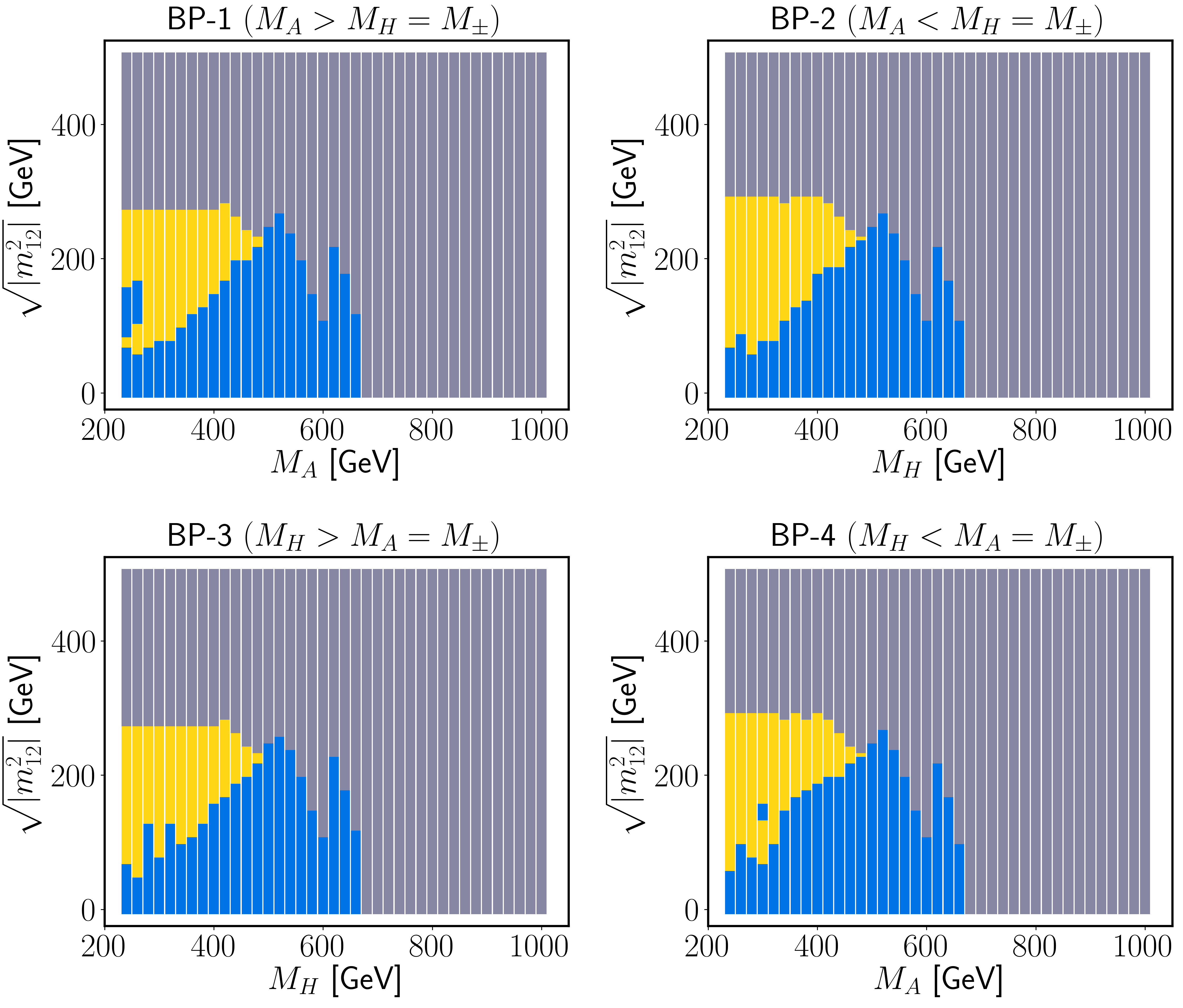}
\caption{
The GM constraints (excluding yellow) combined with the unitarity and BFB constraints (excluding gray) on some benchmarks tabulated in Table~\ref{tab:2HDMexotic}, with $\cos(\beta-\alpha)=0$ and $\tan\beta=1.0$. 
The loop corrections and the $s$ dependence have been included.
The blue regions are the largest allowed regions by all the constraints in the grid scan\,---\,see the text for more details.
}
\label{fig:THDM_constraints_exotictb1}
\end{figure}

For scenario (i), we perform a grid scan of $M_H = M_A = M_\pm$ from $200\,\GeV$ to $1\,\TeV$ at a step of $10\,\GeV$, and $m_{12}^2$ from $0$ to $-(500\,\GeV)^2$ at a step of $(10\,\GeV)^2$.
The 2HDM mixing angles $(\alpha\,,\beta)$ are taken to be consistent with the current LHC constraints on the $125\,\GeV$ Higgs boson signal strengths as shown in Fig.~\ref{fig:THDM_hsig}.
For scenario (ii), we summarize the benchmark models in Table~\ref{tab:2HDMexotic}.
Taking BP-1 for instance, we perform the grid scan of the heaviest Higgs boson mass $M_A$ from $250\,\GeV$ to $1\,\TeV$ at a step of $10\,\GeV$, and the next heavy Higgs boson mass $M_H=M_\pm$ from $130\,\GeV$ up to $M_A-100\,\GeV$.
The soft $\mathbb{Z}_2$  breaking parameter $m_{12}^2$ still takes the negative values from $0$ to $-(500\,\GeV)^2$ at a step of $(10\,\GeV)^2$.
In addition, we take $\cos(\beta-\alpha)=0$ (known as the alignment limit) in this case for simplicity.
For both scenarios, we set $\tan\beta=1.0$
because larger or smaller  $\tan\beta$ will be more stringently constrained by the perturbative unitarity bounds.

In Fig.~\ref{fig:THDM_constraints_degenerate}, we present the GM constraints for scenario (i),  i.e., the mass-degenerate heavy Higgs boson case.
All the samples generated in the above way are first filtered by the unitarity and BFB bounds and then constrained by the GM conditions.
For comparison, we present results for both a simplistic approach
and an improved approach. For the simplistic case, we adopt the conventional
unitarity and BFB bounds which exclude the gray regions, and then use
the tree-level GM constraint to further exclude the yellow regions,
leaving the blue regions that satisfy all the constraints. The improved
case includes the loop corrections and the $s$ dependence (explained
at the end of Sec.~\ref{subsec:GM}), which change the boundaries of the gray
regions to the red curves and the boundaries between the blue and
yellow regions to the black curves. As one can see, the loop-corrected
GM constraints deviate significantly from the tree-level GM constraints,
but both can be violated typically when $|m_{12}|$ is too large.
Due to the significant loop corrections, for the remaining analyses we will adopt the improved constraints
while the tree-level constraints are only used to qualitatively understand
the results.

In Fig.~\ref{fig:THDM_constraints_exotictb1}, we present these joint constraints for the heavy Higgs mass spectrum with exotic decays in the $(M_{A}\,, \sqrt{|m_{12}^2|})$ or $(M_{H}\,, \sqrt{|m_{12}^2|})$ plane, with 2HDM mixing angles of $\cos(\beta-\alpha)=0$ and $\tan\beta=1.0$.
The allowed regions by the GM constraints should be the same for both Type I and Type II models, provided that  the same 2HDM mixing angles are assumed.
The heaviest neutral Higgs boson masses are always labeled as the $x$ axis.
The blue regions represent the largest allowed regions by the grid scan of the next heavy Higgs mass in each benchmark model.


\section{The phenomenology implications: heavy Higgs boson searches at the LHC}
\label{section:Hpair}

In this section, we will discuss the implications of the GM constraint on the LHC phenomenology of the heavy Higgs boson searches in the general 2HDM.
Since we have found that the GM condition is able to further constrain $m_{12}^2$, 
in addition to the unitarity and BFB bounds,
the actually allowed ranges of the Higgs boson self-couplings are further restricted. 
For the cubic Higgs self-couplings in the physical basis, one may check Ref.~\cite{LopezVal:2009qy} for details.
Accordingly, one can expect that the GM condition will be relevant to the SM-like Higgs boson pair productions and other heavy Higgs search limits at the LHC.

\subsection{The heavy $CP$-even Higgs boson decays into SM-like Higgs boson pairs}

\begin{figure}[tb]
\centering
\includegraphics[width=0.45\textwidth]{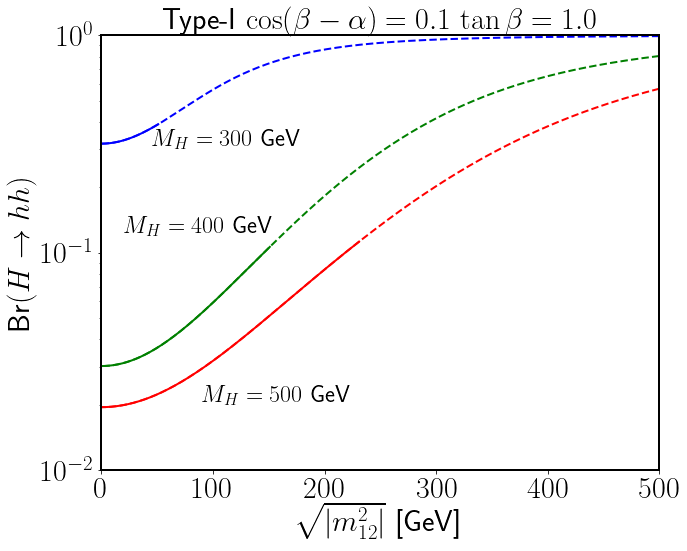}
\includegraphics[width=0.45\textwidth]{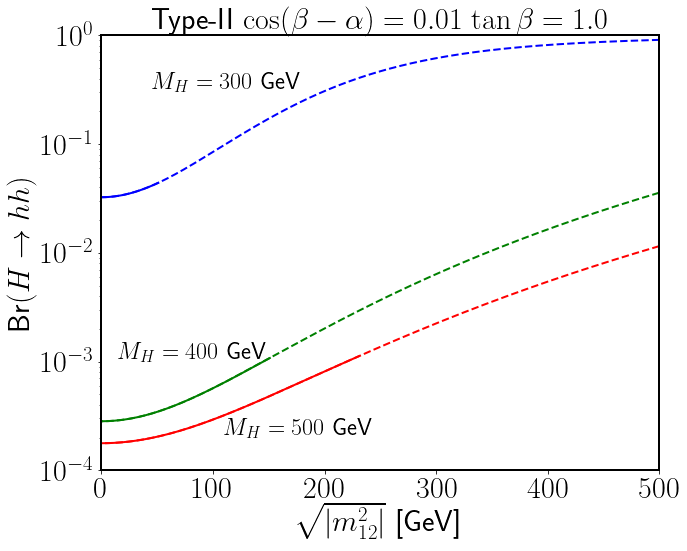}
\caption{
The decay branching fractions of Br$[H\to hh]$, for the Type I model (left panel) and Type II model (right panel).
The solid or dashed curves represent the parameter regions that are allowed or excluded by the GM conditions, respectively. The loop corrections and the $s$ dependence have been included.
}
\label{fig:SMh_BrHtohh}
\end{figure}

We study the resonance productions of the SM-like Higgs boson pair productions for the degenerate heavy Higgs boson scenario.
The exact results for the one-loop Higgs pair production processes at the $pp$ colliders were first studied in Ref.~\cite{Plehn:1996wb}.
For the 2HDM case with nonvanishing inputs of $\cos(\beta-\alpha)$, the leading contribution is due to the heavy $CP$-even Higgs boson resonance $H$.
In the $m_{12}^2<0$ region, we plot the decay branching fraction of Br$[H\to hh]$ for the Type I model (with parameters of $\cos(\beta-\alpha)=0.01$ and $\tan\beta=1.0$) and the Type II model (with parameters of $\cos(\beta-\alpha)=0.1$ and $\tan\beta=1.0$) in Fig.~\ref{fig:SMh_BrHtohh}, for three different inputs of the heavy $CP$-even Higgs boson masses.
The decay branching fractions of Br$[H\to hh]$ are apparently suppressed in the Type II model, with a small $\cos(\beta-\alpha)$ input, as compared to the Type I model.
With the GM condition, the allowed ranges of $\sqrt{ |m_{12}^2| }$ are further restricted, which were also displayed in Fig.~\ref{fig:THDM_constraints_degenerate} previously.

\begin{figure}[tb]
\centering
\includegraphics[width=0.43\textwidth]{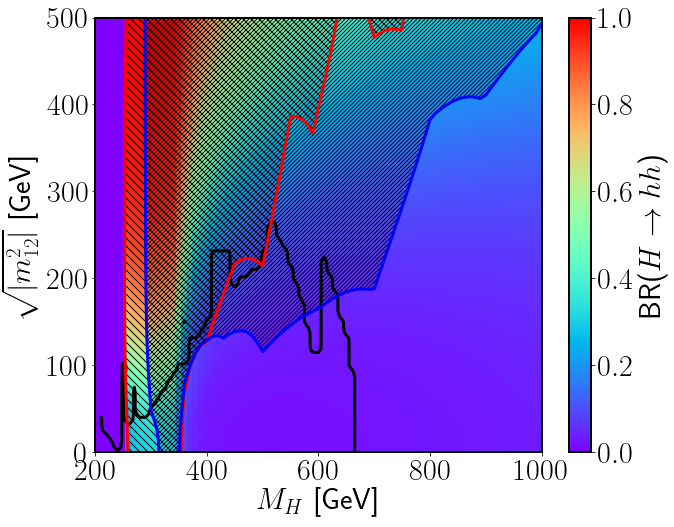}
\hspace{0.04\textwidth} 
\includegraphics[width=0.43\textwidth]{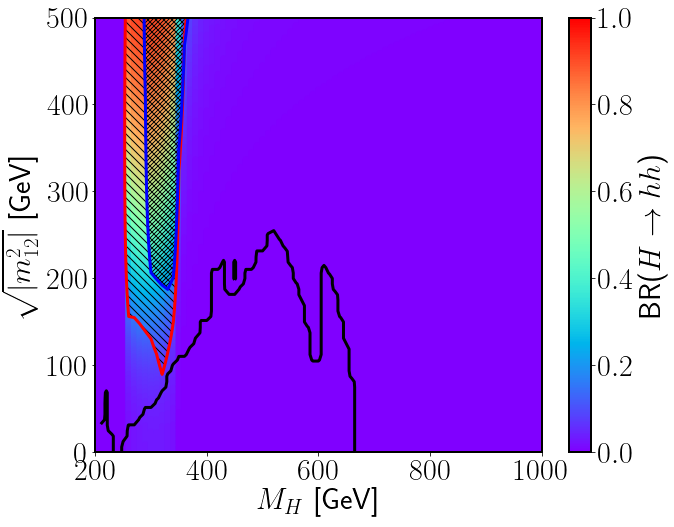}
\caption{
The current LHC $13\,\TeV$ search limits on the resonance productions of SM-like Higgs boson pairs, for the Type I model (left panel) and Type II model (right panel).
The red and blue hatched regions have been excluded by the $hh\to b \bar b \gamma\gamma$ and $hh\to b \bar b b \bar b$, respectively.
The black contours represent the theoretically (including the GM and unitarity and BFB constraints) allowed regions.
}
\label{fig:Htohh_LHC13searches}
\end{figure}

We obtain the heavy $CP$-even Higgs boson production cross sections at the LHC $13\,\TeV$ runs, by using the {\tt Sushi} package~\cite{Harlander:2012pb}.
For the parton distributions, we use NNPDF.
Both the heavy resonance searches for $H\to hh\to b \bar b \gamma\gamma$~\cite{Sirunyan:2018iwt} and $H\to hh \to b \bar b b \bar b$~\cite{Aaboud:2018knk} were taken into account.
In Fig.~\ref{fig:Htohh_LHC13searches}, the current LHC $13\,\TeV$ search limits on the SM-like Higgs boson pairs via these two channels, as well as the theoretically allowed regions, are presented in the $(M_H\,, \sqrt{ |m_{12}^2|})$ plane.
For the Type I model, the current LHC search limits have excluded the heavy Higgs boson mass ranges of $250\,\GeV\lesssim M_H\lesssim 350\,\GeV$.
Meanwhile, the search limits on the Type II model are much smaller in the theoretically allowed region, since the corresponding alignment parameter of $\cos(\beta-\alpha)$ was suppressed from the LHC signal strengths. 
For both Type I and Type II, most of the LHC excluded regions are actually out of the theoretically allowed regions, showing the robustness of  the combinations of GM constraints and other theoretical bounds in the $H\rightarrow hh$ process.

\subsection{The exotic heavy Higgs boson decays}

\begin{figure}[tb]
\centering
\includegraphics[width=0.9\textwidth]{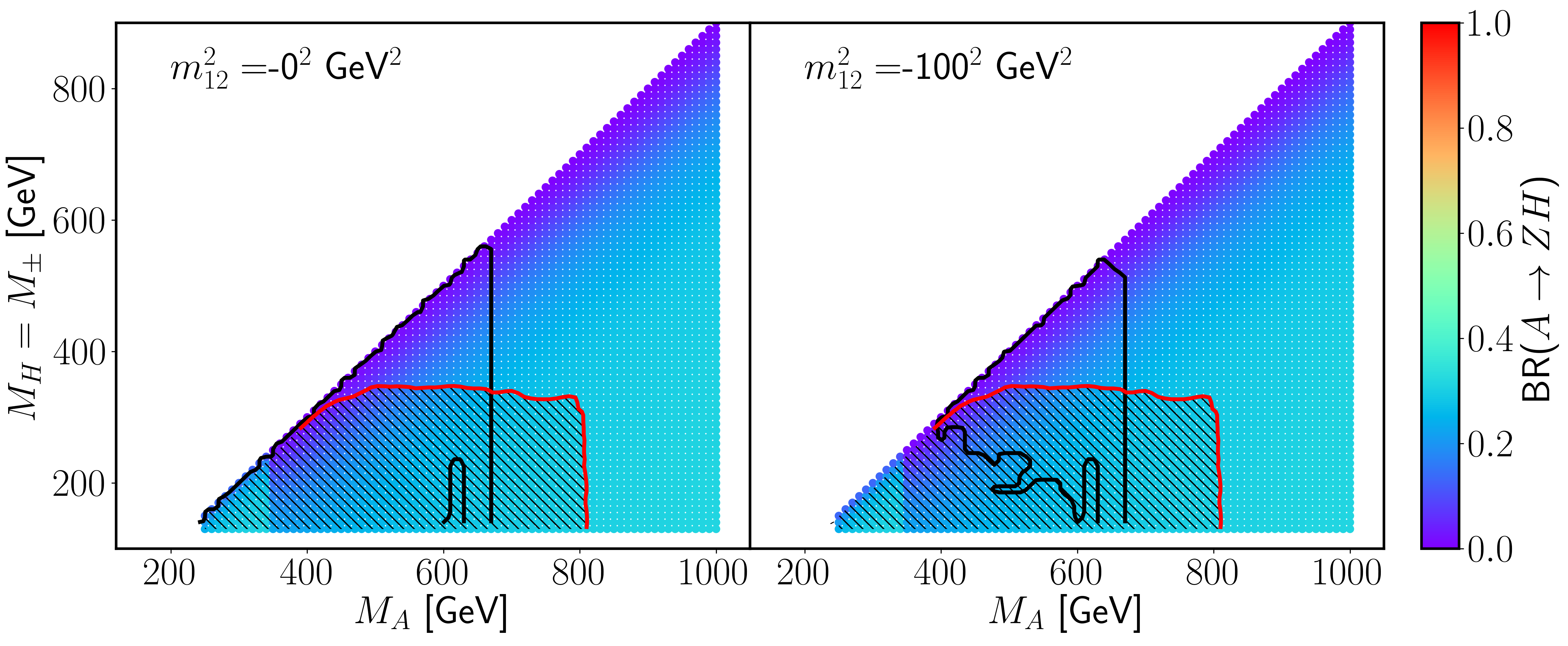}
\caption{
The current LHC $13\,\TeV$ search limits on the $A\to ZH$ in the BP-1 case with $\cos(\beta-\alpha)=0$ and $\tan\beta=1.0$ input.
The red hatched regions have been excluded by the LHC $13\,\TeV$ searches for $A\to ZH$.
The black contours represent the theoretically (including the GM and unitarity and BFB constraints) allowed regions.
}
\label{fig:SSIA_13TeVAZH}
\end{figure}

\begin{figure}[tb]
\centering
\includegraphics[width=0.9\textwidth]{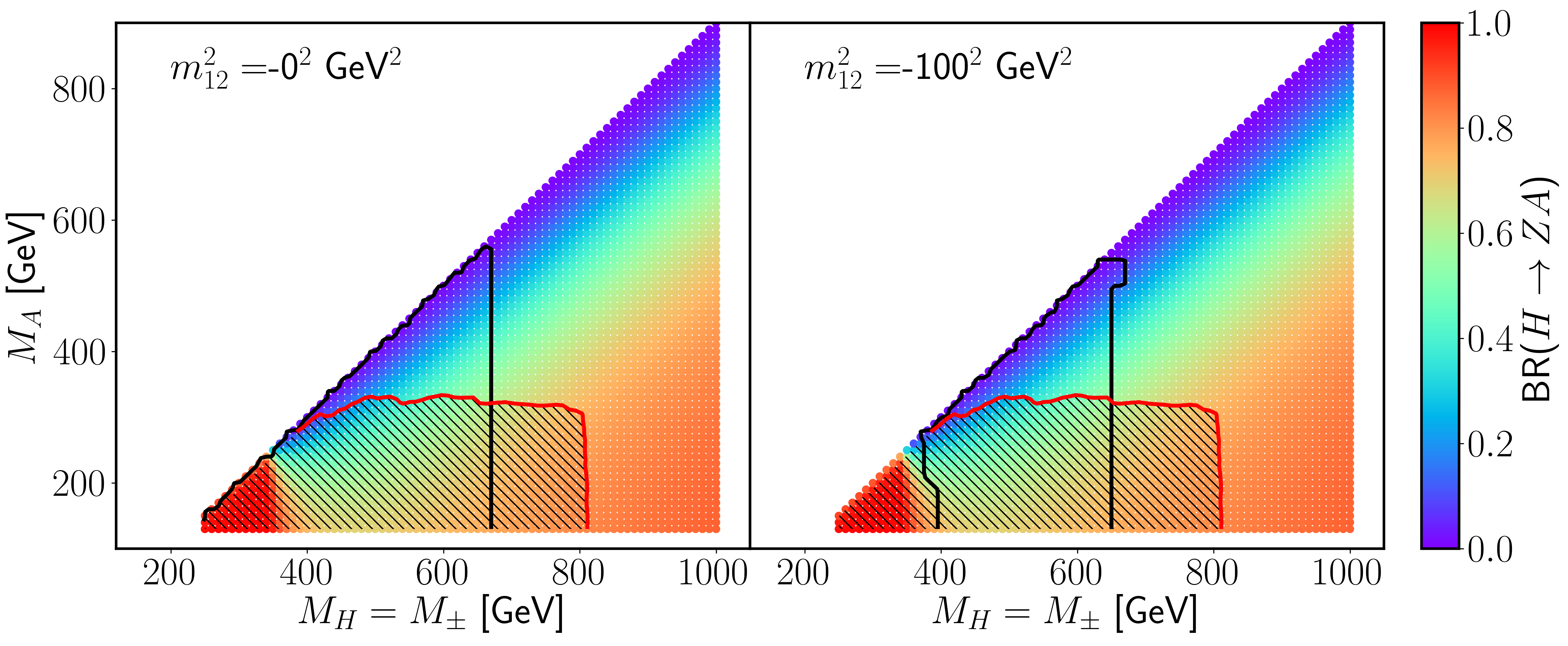}
\caption{
Similar to Fig.~\ref{fig:SSIA_13TeVAZH}, but for the LHC $13\,\TeV$ search limits on $H\to ZA$ in the BP-2 case.
}
\label{fig:SSIB_13TeVHZA}
\end{figure}

\begin{figure}[tb]
\centering
\includegraphics[width=0.9\textwidth]{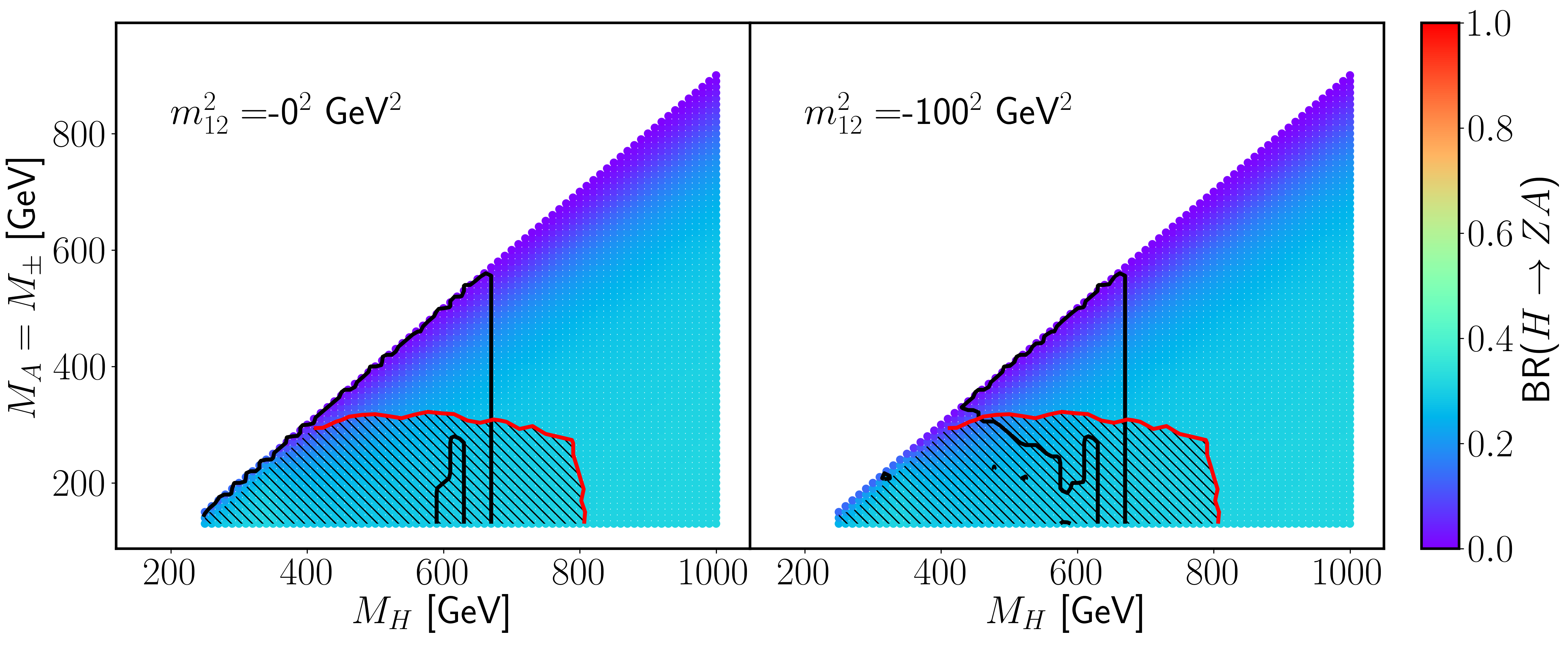}
\caption{
Similar to Fig.~\ref{fig:SSIA_13TeVAZH}, but for the LHC $13\,\TeV$ search limits on $H\to ZA$ in the BP-3 case.
}
\label{fig:SSIIA_13TeVHZA}
\end{figure}

\begin{figure}[tb]
\centering
\includegraphics[width=0.9\textwidth]{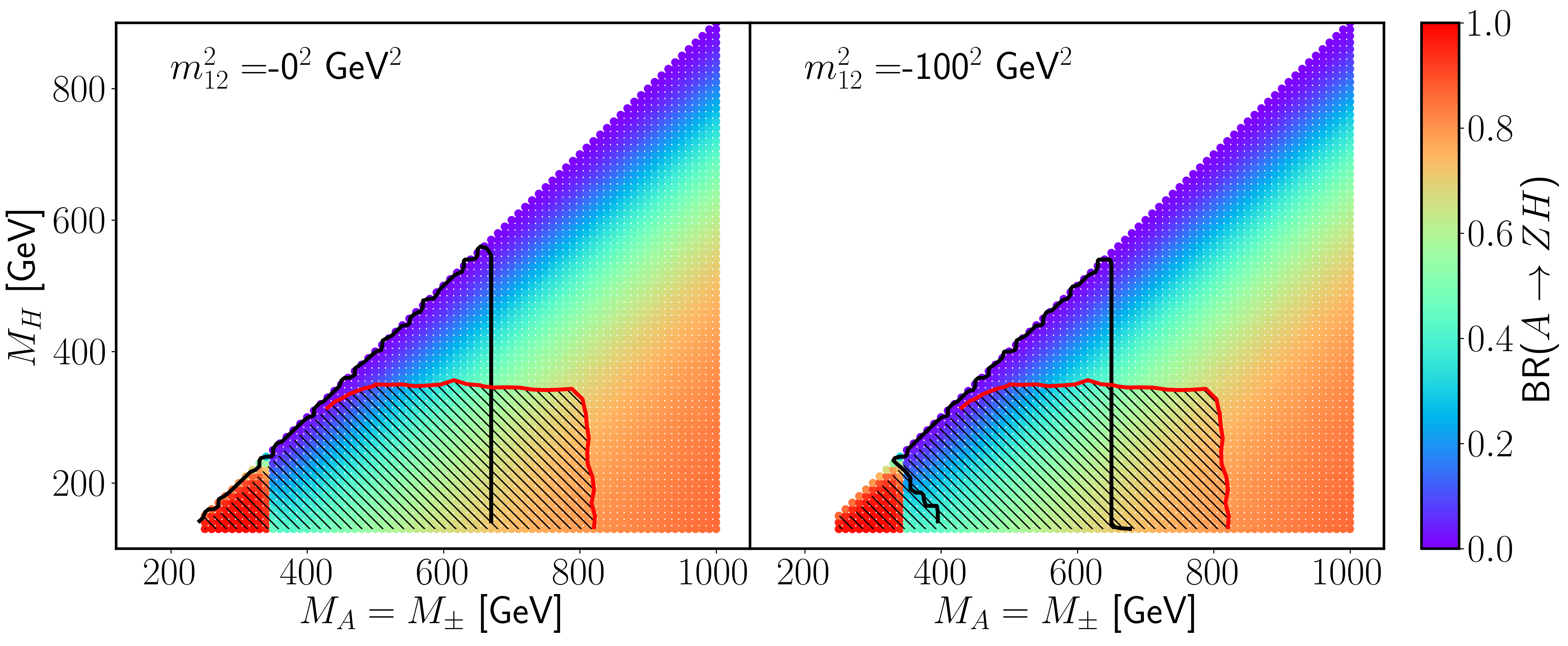}
\caption{
Similar to Fig.~\ref{fig:SSIA_13TeVAZH}, but for the LHC $13\,\TeV$ search limits on $A\to ZH$ in the BP-4 case.
}
\label{fig:SSIIB_13TeVAZH}
\end{figure}

In general, the cubic self-couplings control the partial decay widths of heavy Higgs bosons, such as $H\to hh$, $AA$, $H^+ H^-$, and so on.
The constraints from the GM requirements turn out to be relevant to these partial decay widths, and hence to all possible decay branching fractions of heavy Higgs bosons.
The possible exotic heavy Higgs boson decay modes were previously tabulated in Table~\ref{tab:2HDMexotic} in the alignment limit.

Currently, the most recent LHC search limits on a heavy Higgs boson decaying into a $Z$ boson plus another heavy Higgs boson via the $\ell^+ \ell^- b \bar b$ final state can be found in Ref.~\cite{Aaboud:2018eoy}.
The experimental search limits of $\sigma\times\textrm{Br}[A\to ZH]\times\textrm{Br}[H\to b \bar b]$ have been projected to the two-dimensional plane of $(M_H\,,M_A)$ by assuming that $M_A> M_H$.
Such exotic heavy Higgs decay searches have also been studied in Ref.~\cite{Coleppa:2017lue} via different final states of $2b+4\ell$ or $4b+2\ell$.
Here, we use the observed upper limits on both $gg\to A\to ZH$ and $gg\to H\to ZA$ processes by assuming that the search limits are insensitive to the parity properties of the heavy Higgs bosons.
Through Figs.~\ref{fig:SSIA_13TeVAZH}, \ref{fig:SSIB_13TeVHZA}, \ref{fig:SSIIA_13TeVHZA}, and \ref{fig:SSIIB_13TeVAZH}, we present the current LHC $13\,\TeV$ search limits on the exotic heavy Higgs boson decay modes in the $(M_A\,,M_H)$ plane (for BP-1 and BP-4) or the $(M_H\,,M_A)$ plane (for BP-2 and BP-3).
In Ref.~\cite{Aaboud:2018eoy}, the mass ranges of the experimental searches were taken to be $M_A\in (230\,, 800)\,\GeV$ and $M_H\in (130\,, 700)\,\GeV$.
For all four benchmark planes, the current LHC $13\,\TeV$ experimental search limits have excluded the regions where the next-heaviest Higgs boson masses are $\lesssim 300\,\GeV$.

As shown in Figs.~\ref{fig:SSIA_13TeVAZH}-\ref{fig:SSIIB_13TeVAZH}, for all the cases presented here, the LHC excluded regions partially overlap with  theoretically allowed regions (marked by the black contour), which implies that the GM constraints are complementary to the LHC constraints.
When $m_{12}^2$ changes from zero (left panels) to negative values (right panels), the theoretically allowed regions shrink, leading to preferences for smaller branching ratios. Generally speaking, more negative  $m_{12}^2$ is more stringently constrained by the GM constraints, which is consistent with what has been shown in Figs.~\ref{fig:THDM_constraints_degenerate}-\ref{fig:THDM_constraints_exotictb1}.
This can be understood by the analytical expressions of potential minima in Eqs.~\eqref{eqs:VminBCD}, which can be reduced to 
\beqs
\beqn
V_{\rm min\,,B}&\to&-\frac{(M_h^2 \tan\beta - 2 m_{12}^2 )^2 v^2 \cos^2\beta }{8  \left(M_h^2 \sin^2\beta+M_H^2 \cos^2\beta - m_{12}^2/\tan\beta \right) } \,, \\
V_{\rm min\,,C}&\to&-\frac{( M_h^2 /\tan\beta - 2m_{12}^2 )^2 v^2 \sin^2\beta }{8 \left(M_H^2 \sin^2\beta +M_h^2 \cos^2\beta -m_{12}^2 \tan\beta \right) } \,,\\
V_{\rm min\,,D}&\to&-\frac{v^2}{8} M_h^2 \,.
\eeqn
\eeqs
in the alignment limit of $\cos(\beta-\alpha)=0$.
To have both conditions of $V_{\rm min\,,D}< V_{\rm min\,,B}$ and $V_{\rm min\,,D}< V_{\rm min\,,C}$ hold with a more negative input of $m_{12}^2$, one thus demands a larger input of $M_H$.

For the BP-1 and BP-4 cases, a more negative input of $m_{12}^2$ pushes  
 $M_A$ and $M_H$  closer to each other.
Therefore, one can expect the current experimental searches via the $H\to ZA$ mode become more challenging, since the transverse momenta of final-state $b$ jets and leptons are smaller.
Similar situations can be envisioned for the BP-2 and BP-3 cases as well, but for the different decay mode of $A\to ZH$.
This suggests that the heavy Higgs boson spectrum involving exotic decay modes may be hidden from the LHC experimental searches, with negative inputs of $m_{12}^2$ and the GM constraint taken into account.


\section{Conclusion}
\label{section:conclusion}

In the scalar potential of the general 2HDM, it is likely that several minima may coexist.
The usually considered vacuum can thus become a local minimum, and it may decay into a deeper one.
To avoid this vacuum instability at the tree level, we impose the GM condition to the 2HDM potential.

According to our analysis, it turns out that the GM condition can impose a more stringent bound on the $m_{12}^2$ parameter when it is in the negative region.
Besides, we find that large or small inputs of $\tan\beta$ can impose stringent bounds on the heavy Higgs boson masses for the illustrated cases.
Hence, we focus on the parameter input of $\tan\beta=1.0$ in our discussion.
Two different scenarios in the heavy Higgs boson sector were considered in our analysis.
For the mass-degenerate heavy Higgs bosons, we find that the actually expected decay branching fractions of Br$[H\to hh]$ with a nonvanishing alignment parameter are restricted into smaller ranges.
The current LHC $13\,\TeV$ searches for SM-like Higgs boson pairs via $ b \bar b \gamma\gamma$ and $b \bar b b \bar b$ final states are more sensitive in the Type I benchmark model, compared to the Type II benchmark model with a suppressed $\cos(\beta-\alpha)$ input.
For the heavy Higgs boson spectrum involving exotic decays, the GM constraint can put more stringent bounds with more negative inputs of $m_{12}^2$.
The current LHC $13\,\TeV$ run has performed searches for the heavy Higgs boson with an exotic decay mode of $A\to ZH\to \ell^+ \ell^- b \bar b$.
We projected the experimental search limits on the $(M_A\,,M_H)$ plane (when $M_A>M_H$) or the $(M_H\,,M_A)$ plane (when $M_H>M_A$).
We also note that the GM constraint can squeeze the heavy $CP$-even Higgs boson mass $M_H$ into larger values with the negative inputs of $m_{12}^2$.
Consequently, such parameter regions bring difficulty for the future LHC searches via the exotic heavy Higgs boson decay channels.


\section*{ACKNOWLEDGMENTS}

The work of N.C. is supported by the National Natural Science Foundation of China (under Grant No. 11575176) and Center for Future High Energy Physics (CFHEP). 
The work of Y.C.W. is partially supported by the Natural Sciences and Engineering Research Council of Canada.
N.C. thanks Center for High Energy Physics Peking University for their hospitality when part of this work was prepared.

\bibliographystyle{bibsty}
\bibliography{references}

\end{document}